# Isolated effective coherence (iCoh): causal information flow excluding indirect paths

RD Pascual-Marqui[1,2], RJ Biscay[3], J Bosch-Bayard[4], D Lehmann[1], K Kochi[1], N Yamada[2], T Kinoshita[5], N Sadato[6]

[1]The KEY Institute for Brain-Mind Research, University of Zurich, Switzerland; [2]Department of Psychiatry, Shiga University of Medical Science, Japan; [3]CIMFAV, Universidad de Valparaiso, Chile; [4]Cuban Neuroscience Center, Havana, Cuba; [5]Department of Neuropsychiatry, Kansai Medical University, Japan; [6]Division of Cerebral Integration, National Institute for Physiological Sciences, Okazaki, Japan

Corresponding author:
RD Pascual-Marqui; "pascualm at key.uzh.ch"; "pascualm at belle.shiga-med.ac.jp"; www.uzh.ch/keyinst/loreta.htm
The KEY Institute for Brain-Mind Research, University Hospital of Psychiatry, Zurich, Switzerland
Department of Psychiatry, Shiga University of Medical Sciences, Shiga, Japan

## 1. Abstract

A problem of great interest in real world systems, where multiple time series measurements are available, is the estimation of the intra-system causal relations. For instance, electric cortical signals are used for studying functional connectivity between brain areas, their directionality, the direct or indirect nature of the connections, and the spectral characteristics (e.g. which oscillations are preferentially transmitted). The earliest spectral measure of causality was Akaike's (1968) seminal work on the noise contribution ratio, reflecting direct and indirect connections. Later, a major breakthrough was the partial directed coherence of Baccala and Sameshima (2001) for direct connections. The simple aim of this study consists of two parts: (1) To expose a major problem with the partial directed coherence, where it is shown that it is affected by irrelevant connections to such an extent that it can misrepresent the frequency response, thus defeating the main purpose for which the measure was developed, and (2) To provide a solution to this problem, namely the "isolated effective coherence", which consists of estimating the partial coherence under a multivariate auto-regressive model, followed by setting all irrelevant associations to zero, other than the particular directional association of interest. Simple, realistic, toy examples illustrate the severity of the problem with the partial directed coherence, and the solution achieved by the isolated effective coherence.

For the sake of reproducible research, the software code implementing the methods discussed here (using lazarus free-pascal "www.lazarus.freepascal.org"), including the test data as text files, are freely available at:
https://sites.google.com/site/pascualmarqui/home/icoh-isolated-effective-coherence

## 2. Introduction

Consider a realistic, non-artificial example where time series of local electric potential differences are measured at five sites on the cortex (electrocorticogram, ECoG). An informal description (later to be made more precise using multivariate autoregression) follows. Site 1 has intrinsic activity at 28 Hz, and sends information to Site 2 with a measurable physiological time lag. Site 2 has intrinsic activity at 16 Hz, and sends information to Sites 1, 3, 4, and 5 with a measurable physiological time lag. Sites 3, 4, and 5 have intrinsic independent activities at 23 Hz. Instantaneous information transmission would require ephatic conduction (see e.g. Weiss et al 2013), which is not considered to be present in this realistic example.

In this realistic, non-artificial example, we wish to recover from the time series measurements, all the detailed information about the system: the direct connections, their directionality, and the spectral nature of the information being transmitted.

This example illustrates a very general problem, of realistic nature, that has great interest in the field of human brain function. Other fields of research can as well benefit from the solution to this type of problem.

This type of problem has a higher degree of complexity than what is usually being considered in time series of metabolism (e.g. as measured by fMRI), which have oscillations much lower than 0.1 Hz. It is typical in fMRI to focus only on the directness and directionality of the connections, and not on the fundamental problem encountered in electrophysiology regarding the spectral nature of the information being transmitted. For an example on fMRI connectivity research that lacks spectral information, see Marinazzo et al (2011).

A very narrow and focused review of the literature on methods for estimating direct or indirect connections, the directionality, and their spectral nature, reveals two major contributions:
1. The noise contribution ratio (NCR) of Akaike (1967), which has been extensively used under other names by Saito and Harashima (1981), Kaminski and Blinowska (1991), and Baccala and Sameshima (1998). Details are given below. This method discovers indirect and direct connections without differentiating them, their directionality and spectral characteristics.
2. The partial directed coherence (PDC) of Baccala and Sameshima (2001), which is a measure designed to quantify direct connections that are not confounded by indirect paths, their directionality and spectral characteristics. This is a very widely used measure (cited 650 times at the time of this writing according to "Google-Scholar").

Recently, the PDC has been critically studied by Schelter et al (2009). They pointed out that the normalization used in PDC, i.e. the denominator in the PDC formula (see below) contains all influences from a source node to all other (receiving) nodes, and as a consequence, the PDC decreases in the presence of many nodes, even if the relationship between source and target nodes remains unchanged. The solution to this problem was given in the form of a renormalization of the PDC, using the statistical variance of the strength of the connection.

In this present study, rather the aiming at a re-normalization of the PDC, such as that successfully achieved by Schelter et al (2009), we reformulate the problem from scratch, estimating the partial coherence under a multivariate auto-regressive model, followed by setting all irrelevant associations to zero, other than the particular directional association of interest. This procedure is akin to Pearl's (2000) "surgical intervention" for studying causality. This approach gives the isolated effective coherence (iCoh).

We give a compelling realistic example that shows how the PDC can give incorrect information about the strength of a connection, and incorrect information on its spectral characteristics. And we show how the iCoh solves this problem.

It is also shown how the iCoh can be obtained from Akaike's NCR under the disconnection constraints as used in iCoh. This demonstration does not in any way mean that the iCoh is identical to NCR. Rather it shows that there is a common foundation to both iCoh and NCR, which may aid in interpreting the new and distinct measure: "iCoh".

Finally, it should be mentioned that instantaneous connections can be accommodated in the iCoh measure. As a realistic example of instantaneous connectivity in electrophysiology, consider the use of scalp EEG recordings. These scalp signals should never be used by themselves for studying cortical connectivity (as an example where it is used anyway, see Marinazzo et al 2011), the reason being that the cortical generators do not in general project radially onto the scalp. A "strong connection" between, say, F3 and F4 signals does not in any way imply that there is a strong connection between the underlying left and right frontal cortices. This is explained and illustrated, for instance, in Lehmann et al (2012). Instead, the EEG signals can be used for estimating the cortical activity signals, using a method such as eLORETA (Pascual-Marqui et al, 2011). However, these estimated cortical signals are instantaneously mixed by volume conduction. Formulations for estimating multivariate autoregressive models that take into account this instantaneous mixing which introduces "apparent" instantaneous connectivity can be found, e.g. in Gomez-Herrero et al (2008). Another approach for modeling instantaneous connections consists of formulating a multivariate autoregressive model that includes zero-lag coefficients, as in Faes et al (2013). In this case, no new measures were derived, since all that is needed are the newly estimated lagged coefficients to be plugged into any of the measures, such as PDF or iCoh.

## 3. Multivariate auto-regression, Granger causality, and the cross-spectral density

General background and notation on multivariate autoregressive models, and the corresponding frequency domain spectral density matrix, can be found, for instance, in Akaike (1968) and Yamashita et al (2005).

Consider the stable, stationary multivariate autoregressive model of order *p* written as:

**Eq. 1** $\quad \mathbf{X}(t) = \sum_{k=1}^{p} \mathbf{A}(k)\mathbf{X}(t-k) + \boldsymbol{\varepsilon}(t)$

with $\mathbf{X}(t) \in \mathbb{R}^{q \times 1}$, $q \geq 2$, $\mathbf{A}(k) \in \mathbb{R}^{q \times q}$, $\boldsymbol{\varepsilon}(t) \in \mathbb{R}^{q \times 1}$, and with discrete time $t = 0...N_T - 1$.

Given data sampled in discrete time, the auto-regressive parameters can be estimated by any number of methods, one of which is the simple least squares approach (see e.g. Akaike 1968).

The frequency domain representation is:

**Eq. 2** $\quad \mathbf{X}(\omega) = \mathbf{A}(\omega)\mathbf{X}(\omega) + \boldsymbol{\varepsilon}(\omega)$

where $\mathbf{X}(\omega) \in \mathbb{C}^{q \times 1}$, $\mathbf{A}(\omega) \in \mathbb{C}^{q \times q}$, $\boldsymbol{\varepsilon}(\omega) \in \mathbb{C}^{q \times 1}$ are the discrete Fourier transforms, with discrete frequency $\omega = 0...N_T - 1$. The discrete Fourier transform for obtaining Eq. 2 in practice is described in detail in the appendix.

In this setting, direct Granger causality is defined as follows:
Time series "*j*" directly causes time series "*i*" if the $(i,j)$ element of $\mathbf{A}(\omega)$ is non-zero, i.e.:

**Eq. 3** $\quad \left[\mathbf{A}(\omega)\right]_{ij} \neq 0 \;\Rightarrow\; j \text{ directly Granger causes } i$

see e.g. Granger (1969) and Lutkepohl (2005).

Notation: In what follows, $\left[\mathbf{M}\right]_{ij}$ denotes the element $(i,j)$ of the matrix $\mathbf{M}$.

From Eq. 2, the Hermitian covariance, i.e. the spectral density matrix, is:

**Eq. 4** $\quad \mathbf{S}_x(\omega) = (\mathbf{I} - \mathbf{A}(\omega))^{-1} \mathbf{S}_\varepsilon (\mathbf{I} - \mathbf{A}^*(\omega))^{-1} = (\breve{\mathbf{A}}(\omega))^{-1} \mathbf{S}_\varepsilon (\breve{\mathbf{A}}^*(\omega))^{-1} = \mathbf{B}(\omega) \mathbf{S}_\varepsilon \mathbf{B}^*(\omega)$

where the superscript "*" denotes transpose and complex conjugate, $\mathbf{I}$ is the identity matrix, $\mathbf{S}_\varepsilon \in \mathbb{R}^{q \times q}$ is the noise covariance, and:

**Eq. 5** $\quad \breve{\mathbf{A}}(\omega) = \mathbf{I} - \mathbf{A}(\omega)$

and:

**Eq. 6** $\quad \mathbf{B}(\omega) = (\mathbf{I} - \mathbf{A}(\omega))^{-1} = (\breve{\mathbf{A}}(\omega))^{-1}$

## 4. The inverse of the spectral density matrix and the partial coherence

The inverse of the Hermitian covariance matrix, i.e. the inverse of the spectral density matrix, is:

**Eq. 7** $\quad \mathbf{S}_x^{-1}(\omega) = \breve{\mathbf{A}}^*(\omega) \mathbf{S}_\varepsilon^{-1} \breve{\mathbf{A}}(\omega)$

Its $(i, j)$ element is:

**Eq. 8** $\quad \left[\mathbf{S}_x^{-1}(\omega)\right]_{ij} = \sum_{k=1}^{q} \sum_{l=1}^{q} \left[\breve{\mathbf{A}}^*(\omega)\right]_{ik} \left[\mathbf{S}_\varepsilon^{-1}\right]_{kl} \left[\breve{\mathbf{A}}(\omega)\right]_{lj} = \sum_{k=1}^{q} \sum_{l=1}^{q} \overline{\left[\breve{\mathbf{A}}(\omega)\right]_{ki}} \left[\mathbf{S}_\varepsilon^{-1}\right]_{kl} \left[\breve{\mathbf{A}}(\omega)\right]_{lj}$

where the complex conjugate of a scalar $c$ is denoted as $\bar{c}$. Note that Eq. 8 can be evaluated for subscripts $(i, j)$, $(i, i)$, and $(j, j)$.

The coefficient $\left[\breve{\mathbf{A}}(\omega)\right]_{ij}$ quantifies the effective causal influence for $j \to i$, based on the definition in Eq. 3.

The partial coherence (see e.g. Brillinger 1981) between $(i, j)$ is:

**Eq. 9** $\quad p_{ij}(\omega) = \dfrac{\left[\mathbf{S}_x^{-1}(\omega)\right]_{ij}}{\sqrt{\left[\mathbf{S}_x^{-1}(\omega)\right]_{ii} \left[\mathbf{S}_x^{-1}(\omega)\right]_{jj}}}$

The significance of the partial coherence in a very general setting can be found in Rao (1981). In simple terms, the partial coherence is a measure of association between two complex valued random variables after removing the effect of other measured variables.

## 5. The isolated effective coherence (iCoh) for $j \to i$

The term "effective" as is used here in this context, refers to the direct causal influence of one time series on another, conditional on all other time series, i.e. after removing the effect of all other time series.

The isolated effective coherence (iCoh) for $j \to i$ is defined under the condition that the only non-zero association between the time series is due to $\left[\breve{\mathbf{A}}(\omega)\right]_{ij} \neq 0$. This requires that all other possible associations be set to zero, i.e.:

**Eq. 10** $\quad \left[\mathbf{A}(\omega)\right]_{kl} \equiv 0$, for all $(k, l)$ such that $(k, l) \neq (i, j)$ and $k \neq l$

and:

**Eq. 11** $[\mathbf{S}_\varepsilon]_{kl} \equiv 0$, for all $(k,l)$ such that $k \neq l$

Note that the diagonal elements of $\mathbf{S}_\varepsilon$ and $\mathbf{A}(\omega)$ remain unmodified, since they do not "associate" different nodes.

Emphasis must be placed on the fact that this procedure is meaningful only if the new system with a single association remains stable and stationary.

Plugging Eq. 10 and Eq. 11 into Eq. 8 gives a covariance matrix $\left[\mathbf{S}_x^{-1}(\omega)\right]^{(i \leftarrow j)}$ with elements:

**Eq. 12** $\left[\mathbf{S}_x^{-1}(\omega)\right]_{ij}^{(i \leftarrow j)} = \left[\mathbf{\breve{A}}(\omega)\right]_{ij} \overline{\left[\mathbf{\breve{A}}(\omega)\right]_{ii}} [\mathbf{S}_\varepsilon]_{ii}^{-1}$

**Eq. 13** $\left[\mathbf{S}_x^{-1}(\omega)\right]_{ii}^{(i \leftarrow j)} = [\mathbf{S}_\varepsilon]_{ii}^{-1} \left|\left[\mathbf{\breve{A}}(\omega)\right]_{ii}\right|^2$

**Eq. 14** $\left[\mathbf{S}_x^{-1}(\omega)\right]_{jj}^{(i \leftarrow j)} = [\mathbf{S}_\varepsilon]_{ii}^{-1} \left|\left[\mathbf{\breve{A}}(\omega)\right]_{ij}\right|^2 + [\mathbf{S}_\varepsilon]_{jj}^{-1} \left|\left[\mathbf{\breve{A}}(\omega)\right]_{jj}\right|^2$

where it is assumed that the self-auto-regressions for both nodes are stable, i.e. $\left[\mathbf{\breve{A}}(\omega)\right]_{ii}$ and $\left[\mathbf{\breve{A}}(\omega)\right]_{jj}$ must correspond to stable self-auto-regressions.

Plugging Eq. 12, Eq. 13, and Eq. 14 into Eq. 9 defines the isolated effective coherence (iCoh) for $j \to i$, as the squared modulus of the partial coherence:

**Eq. 15** $\kappa_{i \leftarrow j}(\omega) = \dfrac{\left|\left[\mathbf{\breve{A}}(\omega)\right]_{ij}\right|^2 \left|\left[\mathbf{\breve{A}}(\omega)\right]_{ii}\right|^2 [\mathbf{S}_\varepsilon]_{ii}^{-2}}{[\mathbf{S}_\varepsilon]_{ii}^{-1} \left|\left[\mathbf{\breve{A}}(\omega)\right]_{ii}\right|^2 \left([\mathbf{S}_\varepsilon]_{ii}^{-1} \left|\left[\mathbf{\breve{A}}(\omega)\right]_{ij}\right|^2 + [\mathbf{S}_\varepsilon]_{jj}^{-1} \left|\left[\mathbf{\breve{A}}(\omega)\right]_{jj}\right|^2\right)}$

which clearly satisfies:

**Eq. 16** $0 \leq \kappa_{i \leftarrow j}(\omega) \leq 1$

Note that Eq. 15 is a genuine partial coherence, obtained under certain constraints. The numerator contains an off-diagonal element of the inverse covariance matrix, while the denominator contains the corresponding product of diagonal elements.

For practical computations, Eq. 15 simplifies to:

**Eq. 17** $\kappa_{i \leftarrow j}(\omega) = \dfrac{[\mathbf{S}_\varepsilon]_{ii}^{-1} \left|\left[\mathbf{\breve{A}}(\omega)\right]_{ij}\right|^2}{[\mathbf{S}_\varepsilon]_{ii}^{-1} \left|\left[\mathbf{\breve{A}}(\omega)\right]_{ij}\right|^2 + [\mathbf{S}_\varepsilon]_{jj}^{-1} \left|\left[\mathbf{\breve{A}}(\omega)\right]_{jj}\right|^2}$

The iCoh can be described as the answer to the following question:
"Given a dynamic linear system characterized by its auto-regressive parameters, what would be the equation for partial coherence if all connections are severed, except for the single one of interest?"

Note that the algorithm for computing the iCoh requires:

(1) The estimation of the full, joint, multivariate auto-regressive model only once. No re-estimation is required in the following steps.

(2) Given a pair of nodes and a direction such as $j \to i$, compute Eq. 15 (or equivalently, compute Eq. 17) using the model parameters obtained from step (1).

## 6. Akaike's noise contribution ratio (NCR)

Akaike's (1968) noise contribution ratio (NCR) is based on the spectral representation in Eq. 4, Eq. 5, and Eq. 6. Consider the *i*-th node and consider its univariate spectral density:

**Eq. 18** $$\left[\mathbf{S}_x(\omega)\right]_{ii} = \sum_{k=1}^{q}\sum_{l=1}^{q}\left[\mathbf{B}(\omega)\right]_{ik}\left[\mathbf{S}_\varepsilon\right]_{kl}\left[\mathbf{B}^*(\omega)\right]_{li}$$

Consider the case when the innovations are uncorrelated. Then Eq. 18 simplifies to:

**Eq. 19** $$\left[\mathbf{S}_x(\omega)\right]_{ii} = \sum_{k=1}^{q}\left|\left[\mathbf{B}(\omega)\right]_{ik}\right|^2\left[\mathbf{S}_\varepsilon\right]_{kk}$$

Eq. 19 shows that the spectral power at the node "*i*" receives an additive contribution from the innovation at the *j*-th node $\left[\mathbf{S}_\varepsilon\right]_{jj}$, weighted by the transfer function $\left|\left[\mathbf{B}(\omega)\right]_{ij}\right|^2$. The relative contribution (see e.g. Yamashita et al 2005) the *j*-th node sends to the *i*-th receiving node defines the noise contribution ratio as:

**Eq. 20** $$\gamma_{i \leftarrow j}(\omega) = \frac{\left|\left[\mathbf{B}(\omega)\right]_{ij}\right|^2\left[\mathbf{S}_\varepsilon\right]_{jj}}{\sum_{k=1}^{q}\left|\left[\mathbf{B}(\omega)\right]_{ik}\right|^2\left[\mathbf{S}_\varepsilon\right]_{kk}}$$

which clearly satisfies:

**Eq. 21** $$0 \leq \gamma_{i \leftarrow j}(\omega) \leq 1$$

Note that the NCR is non-zero if $\left[\mathbf{B}(\omega)\right]_{ij}$ is non-zero, and that this can happen even if the direct causality is zero, i.e. even if $\left[\mathbf{A}(\omega)\right]_{ij}$ is zero. Thus, the NCR is a frequency domain measure of total causality, which includes direct and indirect connections.

## 7. Other measures published after 1968 that are equivalent to Akaike's NCR

Note that a number of frequency domain measures of causality published very much after Akaike's 1968 NCR are actually equivalent to the NCR as is, or are equivalent to the NCR under some simple constraint such as an identity matrix for the innovation covariance.

This assertion can be checked and verified by the interested reader.

In this sense, it is most unfortunate that Akaike's seminal contribution hasn't been properly acknowledged as predating by many years the methods of Saito and Harashima (1981), Kaminski and Blinowska (1991), and Baccala and Sameshima (1998).

## 8. Constraining the NCR and equivalence to the iCoh

As mentioned in Yamashita et al (2005), the NCR in Eq. 20 can be applied in an ad hoc manner for the case of correlated innovations, by simply setting the off-diagonal elements of the innovation covariance matrix $\mathbf{S}_\varepsilon$ to zero, which is equivalent to enforcing Eq. 11.

Now consider the case when both constraints given by Eq. 10 and Eq. 11 are forced (or used in an ad hoc manner) on the definition for the NCR. As with the iCoh, this corresponds to the condition that the only non-zero association between time series is due to $\left[\breve{\mathbf{A}}(\omega)\right]_{ij} \neq 0$. The matrix $\mathbf{B}(\omega)$ in Eq. 6 can be explicitly computed due to its simple structure, which has as non-zero elements all diagonal elements $\left[\breve{\mathbf{A}}(\omega)\right]_{kk} \neq 0$, and the element $\left[\breve{\mathbf{A}}(\omega)\right]_{ij} \neq 0$.

It can then by easily shown that the constrained NCR and the iCoh are identical.

Informally, the constrained NCR can be described as the answer to the following question: "Given a dynamic linear system characterized by its auto-regressive parameters, what would be the equation for NCR if all connections are severed, except for the single one of interest?"

## 9. The partial directed coherence (PDC) and the generalized partial directed coherence (gPDC)

These definitions are replicated here for the sake of completeness.

The PDC is:

**Eq. 22**
$$\left|\bar{\pi}_{ij}(\omega)\right|^2 = \frac{\left|\left[\breve{\mathbf{A}}(\omega)\right]_{ij}\right|^2}{\left[\breve{\mathbf{A}}^*(\omega)\breve{\mathbf{A}}(\omega)\right]_{jj}} = \frac{\left|\left[\breve{\mathbf{A}}(\omega)\right]_{ij}\right|^2}{\sum_{k=1}^{q}\left|\left[\breve{\mathbf{A}}(\omega)\right]_{kj}\right|^2}$$

which corresponds to Baccala and Sameshima (2001), equation #18 therein.

The gPDC is:

**Eq. 23**
$$\left|\pi_{ij}^w(\omega)\right|^2 = \frac{[\mathbf{S}_\varepsilon]_{ii}^{-1}\left|\left[\breve{\mathbf{A}}(\omega)\right]_{ij}\right|^2}{\left[\breve{\mathbf{A}}^*(\omega)(diag\mathbf{S}_\varepsilon)^{-1}\breve{\mathbf{A}}(\omega)\right]_{jj}} = \frac{[\mathbf{S}_\varepsilon]_{ii}^{-1}\left|\left[\breve{\mathbf{A}}(\omega)\right]_{ij}\right|^2}{\sum_{k=1}^{q}[\mathbf{S}_\varepsilon]_{kk}^{-1}\left|\left[\breve{\mathbf{A}}(\omega)\right]_{kj}\right|^2}$$

which corresponds to Baccala et al (2007), equation #11 therein. In Eq. 23, the notation "$diag\mathbf{M}$" denotes a diagonal matrix formed by the diagonal elements of the $\mathbf{M}$.

## 10. The PDC and gPDC are neither coherences nor partial coherences

It is important to note that neither the partial directed *coherence*, nor the generalized partial directed *coherence* are *coherences* in any sense. Note that both satisfy:

**Eq. 24** $$\sum_{j=1}^{q}\left|\pi_{ij}^{w}(\omega)\right|^{2} = \sum_{j=1}^{q}\left|\overline{\pi}_{ij}(\omega)\right|^{2} = 1$$

which is a property not related to nor satisfied by a coherence matrix.

It is important to clarify this fact, because even though the PDCs certainly do give frequency information on the direct connections, they do so in such a way which is fundamentally different from a proper, genuine coherence.

## 11. Toy Examples

Two toy examples will be considered. In both cases, 25600 time samples were generated (after discarding the first 1000 time samples) and used for all estimation procedures. Reported frequency values in Hz units correspond to the assumption that the sampling rate is 256 Hz. All frequency domain measures are shown from 1 to 127 Hz.

Multivariate auto-regressive models are estimated from the data by common least squares, assuming an auto-regressive order $p=3$, although both toy examples are of actual auto-regressive order $p=2$.

### Toy Example 9.1.

Toy Example 9.1. is taken from Baccala and Sameshima (2001), corresponding to their example #5, shown in their figure #4. The Baccala and Sameshima (2001) paper already compares their PDC measure with Akaike's NCR, which is referenced therein as the directed transfer function (DTF) of Kaminski and Blinowska (1991). For this reason, the NCR is not shown in this present study.

Figure 1 is a schematic representation of the direct connections among 5 nodes.

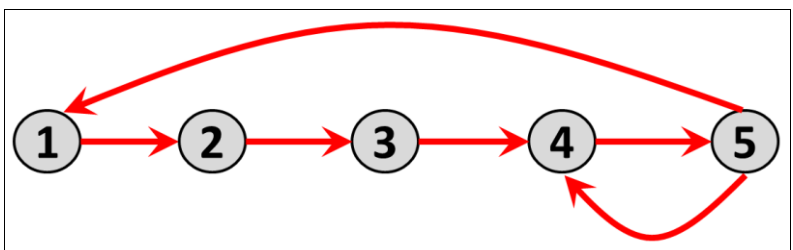

Figure 1: Toy Example 9.1. Schematic representation of the direct wiring among 5 nodes, in example#5 from Baccala and Sameshima (2001).

Table 1 shows the time domain auto-regressive parameters.

| | | | | | |
|---|---|---|---|---|---|
| $\mathbf{A}(1) =$ | 1.3435 | 0 | 0 | 0 | 0 |
| | -0.5 | 0 | 0 | 0 | 0 |
| | 0 | 0 | 0 | 0 | 0 |
| | 0 | 0 | -0.5 | 0.3536 | 0.3536 |
| | 0 | 0 | 0 | -0.3536 | 0.3536 |
| | | | | | |
| $\mathbf{A}(2) =$ | -0.9025 | 0 | 0 | 0 | 0.5 |
| | 0 | 0 | 0 | 0 | 0 |
| | 0 | 0.4 | 0 | 0 | 0 |
| | 0 | 0 | 0 | 0 | 0 |
| | 0 | 0 | 0 | 0 | 0 |
| | | | | | |
| $diagonal\ \mathbf{S}_\varepsilon =$ | 1 | 1 | 1 | 1 | 1 |

Table 1: Toy Example 9.1. Time domain auto-regressive parameters for 5 nodes, in example#5 from Baccala and Sameshima (2001).

Figure 2 displays 1024 time samples from the time series.

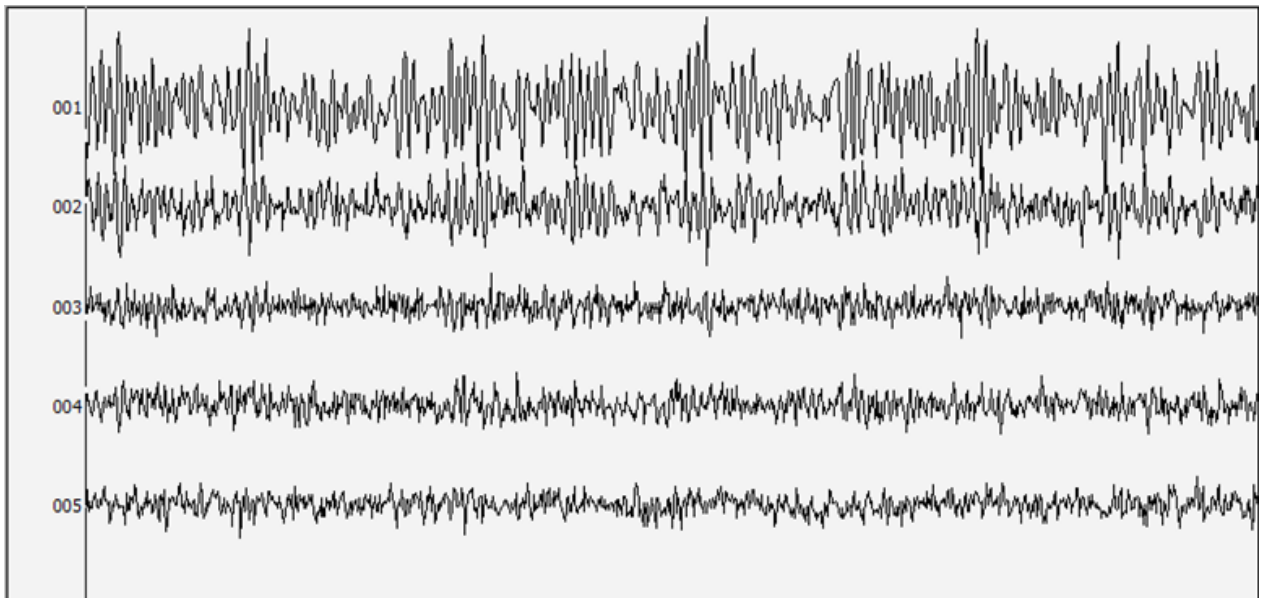

Figure 2: Toy Example 9.1. Time series display of 1024 time samples.

Figure 3 shows the coherence and the spectra.

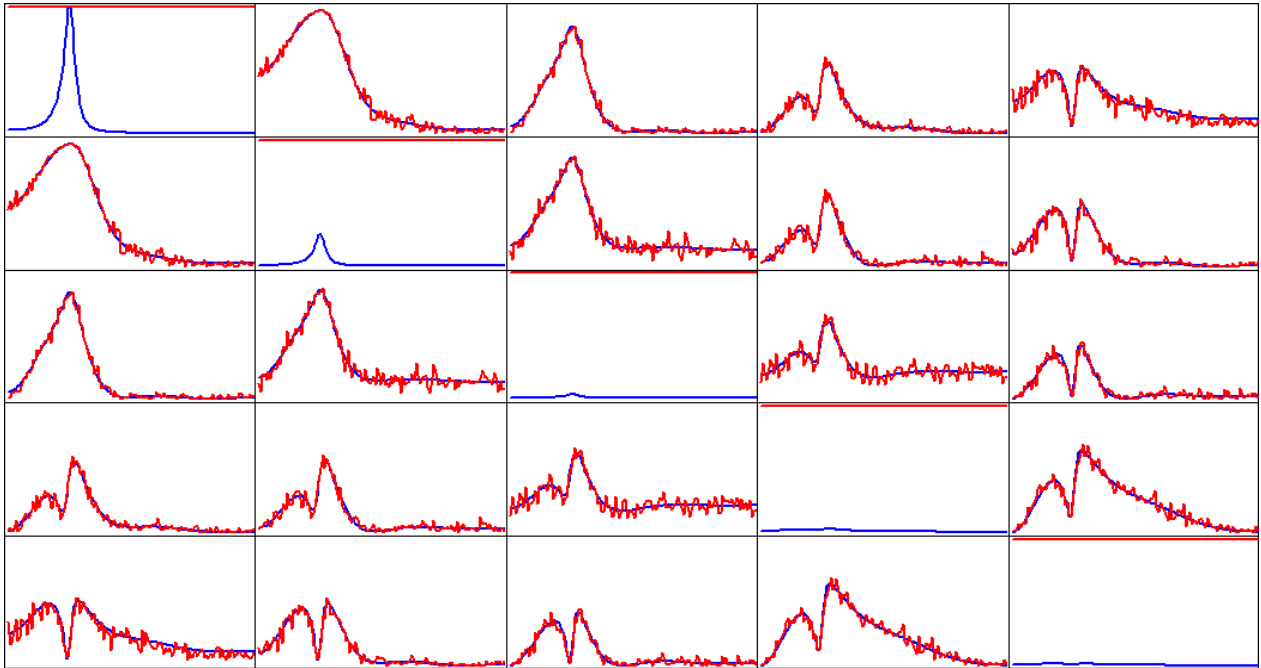

Figure 3: Toy Example 9.1. Auto-regressive spectra (diagonal, scaled to unit maximum power) and squared modulus of the coherence. Vertical axis: 0 to 1. Frequency axis: 1 to 127 Hz. Auto-regression based coherence estimates are shown in blue, while periodogram based estimates are shown in red. Spectral peak at 33 Hz (e.g. row 1, column 1). Coherence peaks at 22 and 35 Hz (e.g. row 4, column 1).

Figure 4 shows the iCoh (Eq. 15) and the gPDC (Eq. 23). Note that in this simple toy example taken from Baccala and Semashima (2001), the two methods give very similar results. These results show that the only non-zero values correctly detect the direction of the directly connected nodes in Figure 1. However, note that iCoh is slightly larger than gPDC for the connections of from node #5 to nodes #1 and #4.

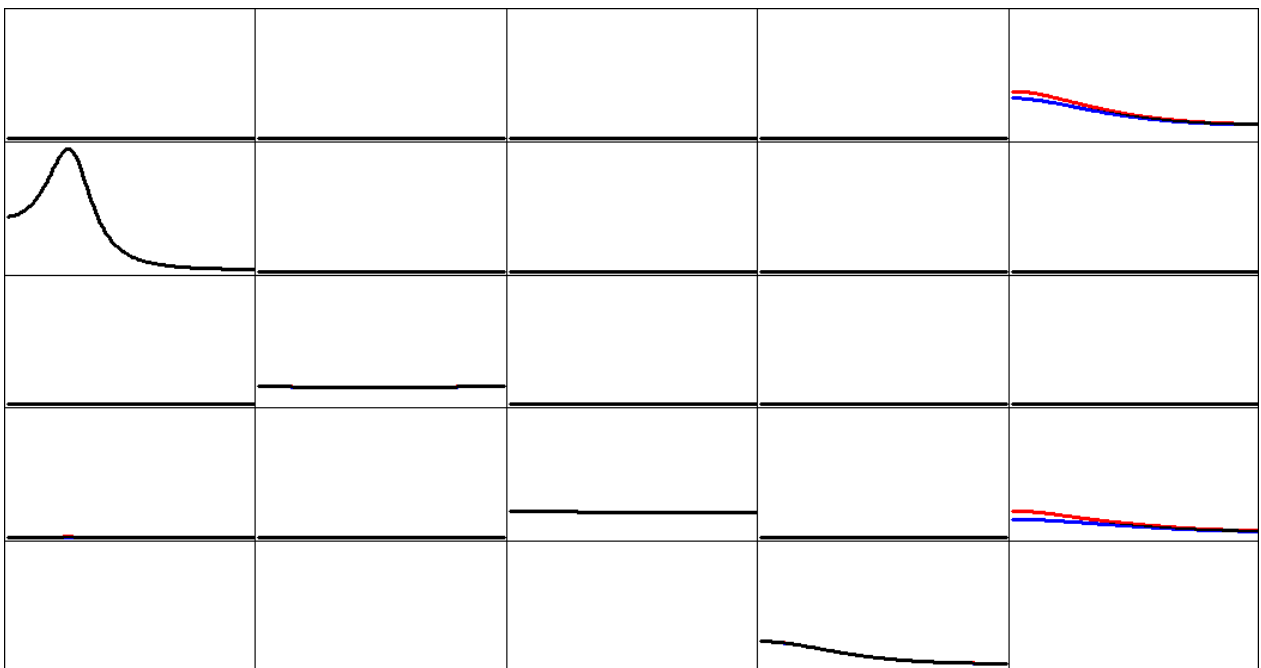

Figure 4: Toy Example 9.1. Isolated effective coherence (iCoh, Eq. 15) shown in RED, and the generalized partial directed coherence (gPDC, Eq. 23) shown in BLUE. Overlap of both curves is shown in BLACK. Vertical axis: 0 to 1. Frequency axis: 1 to 127 Hz. Columns are senders, rows are receivers. Coherence peak at 33 Hz (row 2, column 1).

The fact that the two methods (iCoh and gPDC) give similar results is only due to the simplicity of this example. As will be shown in the next toy example, the two methods can give very different results.

### Toy Example 9.2.

Figure 1 is a schematic representation of the direct connections among 5 nodes for Toy Example 9.2.

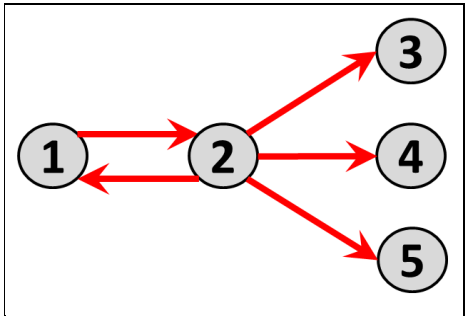

Figure 5: Toy Example 9.2. Schematic representation of the direct wiring among 5 nodes.

Table 2 shows the time domain auto-regressive parameters for Toy Example 9.2.

| | | | | | |
|---|---|---|---|---|---|
| | 1.5 | -0.25 | 0 | 0 | 0 |
| | -0.2 | 1.8 | 0 | 0 | 0 |
| $\mathbf{A}(1)=$ | 0 | 0.9 | 1.65 | 0 | 0 |
| | 0 | 0.9 | 0 | 1.65 | 0 |
| | 0 | 0.9 | 0 | 0 | 1.65 |
| | | | | | |
| | -0.95 | 0 | 0 | 0 | 0 |
| | 0 | -0.96 | 0 | 0 | 0 |
| $\mathbf{A}(2)=$ | 0 | -0.8 | -0.95 | 0 | 0 |
| | 0 | -0.8 | 0 | -0.95 | 0 |
| | 0 | -0.8 | 0 | 0 | -0.95 |
| | | | | | |
| *diagonal* $\mathbf{S}_\varepsilon =$ | 1 | 1 | 1 | 1 | 1 |

Table 2: Toy Example 9.2. Time domain auto-regressive parameters for 5 nodes.

Figure 6 displays 1024 time samples from the time series.

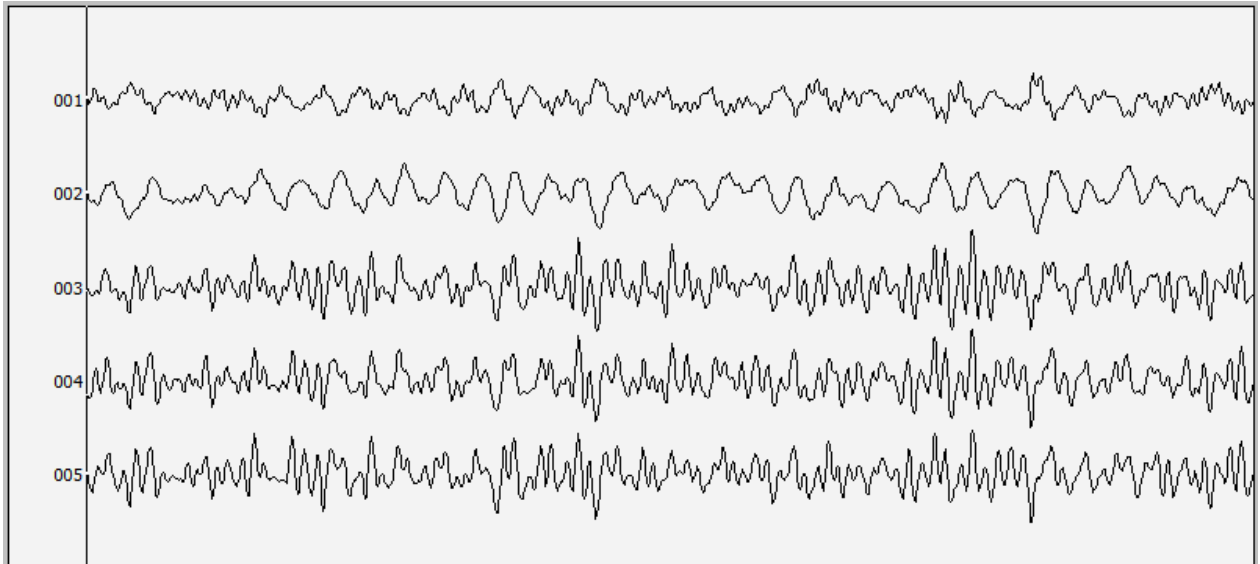

Figure 6: Toy Example 9.2. Time series display of 1024 time samples.

Figure 7 shows the coherence and the spectra. Note that practically all coherences reach very high values (close to 1) at some frequency.

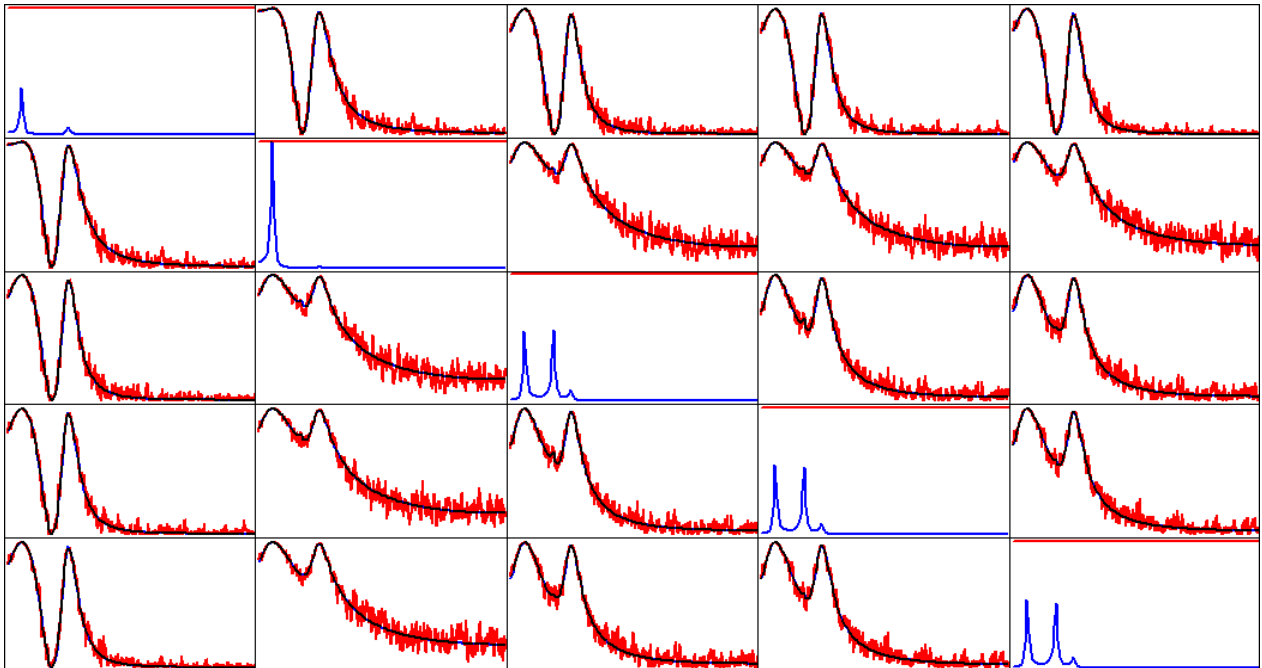

Figure 7: Toy Example 9.2. Auto-regressive spectra (diagonal, scaled to unit maximum power) and squared modulus of the coherence. Vertical axis: 0 to 1. Frequency axis: 1 to 127 Hz. Auto-regression based coherence estimates are shown in blue, while periodogram based estimates are shown in red. Spectral peaks at 8, and 32 Hz present for all diagonals; additional peak at 23 Hz for last three diagonals. Coherence peaks at 8 and 32 Hz.

Figure 8 shows the iCoh (Eq. 15) and the gPDC (Eq. 23). Note that in this toy example, the two methods give very different results with respect to node #2 as sender (column 2).

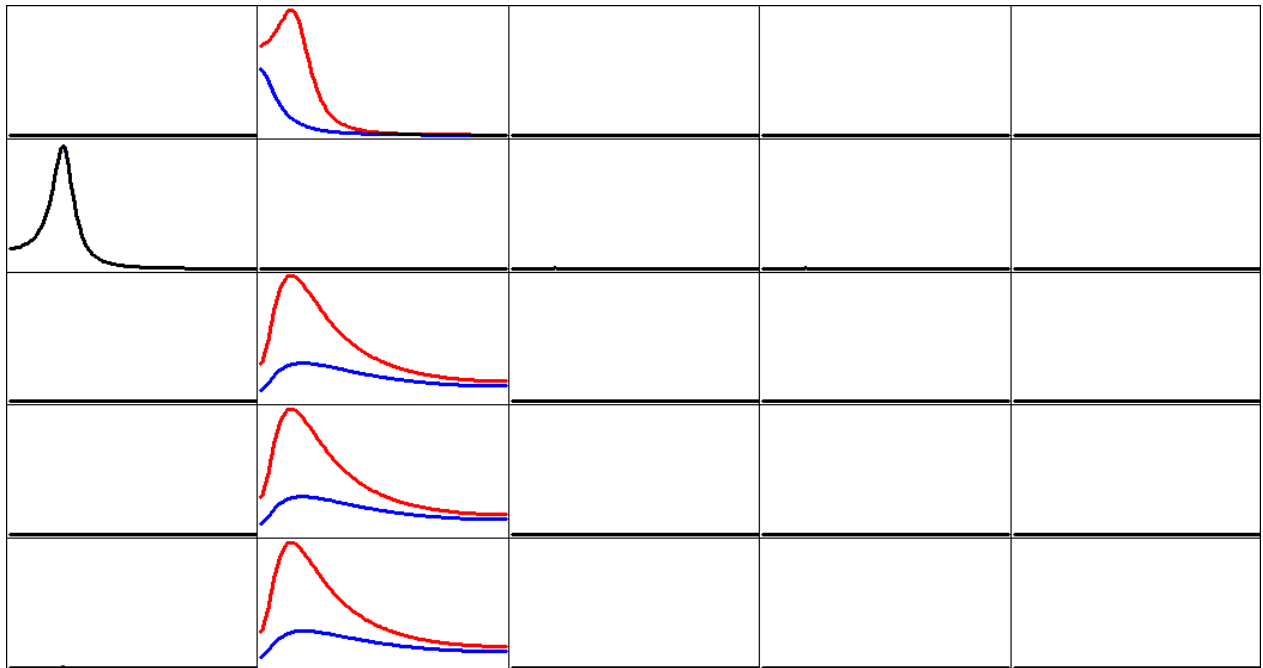

Figure 8: Toy Example 9.2. Isolated effective coherence (iCoh, Eq. 15) shown in RED, and the generalized partial directed coherence (gPDC, Eq. 23) shown in BLUE. Overlap of both curves is shown in BLACK. Vertical axis: 0 to 1. Frequency axis: 1 to 127 Hz. Columns are senders, rows are receivers. Coherence peak in column 2 occurs at 28 Hz. Coherence peak for iCoh in column 2 occurs at 16 Hz. Coherence peak for gPDC in column 2, row 1 occurs at 1 Hz; and Coherence peak for gPDC in column 2, rows 3, 4, and 5 occur at 23 Hz.

## 12. Discusion: the good and the bad of iCoh and gPDC

In general, both iCoh and gPDC give a frequency representation of direct, directional information flow, which is in general theoretical agreement with Granger causality.

However, due to the denominator in the definition of gPDC, this measure can produce very misleading results.

By definition and construction of the auto-regressive coefficients in Table 2, the intrinsic, isolated frequency generated by node #2 is 16 Hz. However, as shown in Toy Example 9.2., the gPDC would lead to the conclusion that node #2 is sending a small amount of low frequency information to node #1 (spectral peak at 1 Hz). This conclusion is incorrect and misleading.

In contrast, iCoh correctly identifies not only the peak frequency of the information being transmitted from node #1 to node #2, but also the strength of the connection, which is, by construction, very high.

The reason for this undesirable behavior of the gPDC is its denominator. When computing the gPDC for $j \rightarrow i$, the denominator includes many terms that contain irrelevant information, such as the coefficients corresponding to node "$j$" as a sender to all other non-related and irrelevant nodes.

The problem with gPDC is actually worse, in that not only misleading frequency characteristics can be produced, but also in that the actual values can be incorrectly low. This is also clear from Toy Example 9.2., where the classical coherences between (2,3), (2,4), and (2,5) are high (reaching values near to 1 at some frequency, see Figure 7), whereas the gPDC reports maximum value lower than 0.5. In contrast, the iCoh shows correctly higher values, in agreement with the classical coherence.

These results cast doubt on the use of the gPDC in practice (see e.g. Sato et al 2009), since it is possible (but not necessarily always so) that it reports incorrectly both the frequency characteristics of the information being transferred, and the actual strength of the association.

The isolated effective coherence (iCoh) does not share these problems. Furthermore, it is a genuine coherence under certain constraints, unlike the gPDC which is not a proper coherence.

One important feature of iCoh is that it does not show the frequency characteristics that are actually being transmitted, which correspond to 8 and 32 Hz, as shown in Toy Example 9.2., the spectral peaks in Figure 7. By definition and construction of the AR coefficients (Table 2), the intrinsic, isolated frequencies of node #1 and #2 are 28 Hz and 16 Hz, respectively. And it is these isolated, intrinsic frequencies which are correctly detected by iCoh, together with the quantitative strength of the connection.

However, it would be desirable to also have a measure of effective coherence that at the same time provides correct information on: (1) actual frequency characteristics being transmitted, (2) directedness of the connections, and (3) direction of the connections. At the moment of this writing, work is in progress in this direction.

## 14. Appendix: The discrete Fourier transform

Let $z(t)$, with discrete time $t=0...N_T-1$, denote a time series. Its discrete Fourier transform is defined as:

**Eq. 25**
$$z(\omega) = \sum_{t=0}^{N_T-1} z(t) \exp\left(-\iota \frac{2\pi t \omega}{N_T}\right)$$

with $\iota = \sqrt{-1}$, and with discrete frequency $\omega = 0...N_T-1$.

In the particular case of the discrete Fourier transform of autoregressive coefficients $b(k)$ defined for $k=1...p$, the coefficients are redefined for $k=0...N_T-1$, with:

$b(k) = 0$ for $k=0$ and $k>p$

The discrete Fourier transform is calculated for this new series, using Eq. 25, giving $b(\omega)$ for all discrete frequencies $\omega = 0...N_T-1$.

## 15. Acknowledgment

RJ Biscay was partly support by the CONICYT ACT 1112 and CONACYT 131771 projects.